\documentclass[useAMS,usenatbib]{mn2e}
\usepackage{graphicx}
\usepackage{psfig}

\title{Absolute Dimensions and Apsidal Motion of the Eccentric Binary V731~Cephei}
\author[V. Bak{\i}\c{s} et al.]
{V. Bak{\i}\c{s}$^1$\thanks{bakisv@comu.edu.tr}, M. Zejda$^2$, \.{I}. Bulut$^1$, M. Wolf$^3$, S. Bilir$^4$, H. Bak{\i}\c{s}$^1$, O.
Demircan$^1$,
\newauthor
J. W. Lee$^5$, M. \v{S}lechta$^6$, B. Ku\v{c}erov\'a$^2$ \\ \\
{\small{$^1$ \c{C}anakkale Onsekiz Mart Univ. Observatory, Terzio\v{g}lu Campus, TR-17040, \c{C}anakkale, Turkey}} \\
{\small{$^2$ Institute of Theoretical Physics and Astrophysics,
Masaryk University, Kotl\'a\v{r}sk\'a 2, CZ-611~37~Brno, Czech Republic}} \\
{\small{$^3$ Astronomical Institute, Faculty of Mathematics and Physics, Charles University Prague, V Hole\v{s}ovi\v{c}k\'ach 2,}} \\
{\small{CZ-182~00~Praha~8, Czech Republic}} \\
{\small{$^4$Istanbul University Science Faculty, Department of Astronomy and Space Sciences, 34119, University-Istanbul, Turkey}}\\
{\small{$^5$ Korea Astronomy and Space Science Institute, Daejeon 305-348, Korea}} \\
{\small{$^6$ Astronomical Institute, Academy of Sciences, CZ-251~65~Ond\v{r}ejov, Czech Republic}}}

\begin{document}

\pagerange{\pageref{firstpage}--\pageref{lastpage}} \pubyear{2008}

\maketitle

\label{firstpage}

\begin{abstract}

A detailed analysis of new and existing photometric, spectroscopic and spatial distribution data of the eccentric binary V731~Cep was performed.
Spectroscopic orbital elements of the system were obtained by means of cross-correlation technique. According to the solution of radial
velocities with {\em UBVR$_c$} and {\em I$_c$} light curves, V731~Cep consists of two main-sequence stars with masses M$_{1}$=2.577 (0.098)
M$_{\odot}$, M$_{2}$=2.017 (0.084) M$_{\odot}$, radii R$_{1}$=1.823 (0.030) R$_{\odot}$, R$_{2}$=1.717 (0.025) R$_{\odot}$, and temperatures
T$_{eff1}$=10700 (200) K, T$_{eff2}$=9265 (220) K separated from each other by \textit{a}=23.27 (0.29) R$_{\odot}$ in an orbit with inclination
of 88$^{\circ}$.70 (0.03). Analysis of the O--C residuals yielded a rather long apsidal motion period of $U$$=$10000(2500) yr compared to the
observational history of the system. The relativistic contribution to the observed rates of apsidal motion for V731~Cep is significant (76 per
cent). The combination of the absolute dimensions and the apsidal motion properties of the system yielded consistent observed internal structure
parameter (log$\bar{k}_{2,obs}$ = $-$2.36) compared to the theory (log$\bar{k}_{2,theo}$ = $-$2.32). Evolutionary investigation of the binary by
two methods (Bayesian and evolutionary tracks) shows that the system is $t=$133(26) Myr old and has a metallicity of $[M/H]=-0.04(0.02)$ dex.
The similarities in the spatial distribution and evolutionary properties of V731~Cep with the nearby ($\rho\sim$3$^{\circ}$.9) open cluster NGC
7762 suggests that V731~Cep could have been evaporated from NGC~7762.

\end{abstract}


\section{Introduction}

Study of eclipsing binaries is still the most effective way of determining the absolute parameters of stars, especially from the spectroscopic
and photometric analysis of detached double-lined eclipsing binaries, masses and radii of the components can be obtained with a precision of
$\sim$1 per cent (e.g. Southworth et al. 2005; Bak\i\c{s} et al. 2008), the limit precision for stellar evolutionary tests (Andersen 1991).
Among the stars, those in the upper main sequence band are particularly useful in empirical tests of the convection formulae used in various
evolution codes, while systems containing unevolved stars are useful in testing opacity and metallicity effects in near-ZAMS (Zero Age Main
Sequence) models. V731~Cep, the binary discussed in the present paper, is of the latter type.

The variability of V731~Cep (GSC~4288~0168; brightness at maximum V$\sim$10.5 mag; orbital period $P$$\sim$6.06 d) was discovered by Bak\i\c{s}
et al. (2003) (hereafter B03). A brief history of V731~Cep was given by Bak\i\c{s} et al. (2007) (hereafter B07) who presented photometric light
curves in {\em BVR$_c$} and {\em I$_c$} bands and limited spectroscopic observations. Within the scope of a
project to study close eclipsing binaries of SB2 type, we were able to obtain new times of minima as well as new spectroscopic data for V731~Cep.

The system has an eccentric orbit ($e=0.0165$), which makes it an important astrophysical tool for the investigation of internal structures of
its components, if the rotation period of the apsides is precisely determined. The apsidal motion period together with system geometry allow the
computation of the observed internal structure constant (ISC) to be compared with theoretical internal structure computations. The primary
motivation for the present paper is to obtain the parameters of the close binary system V731~Cep, to discuss the evolutionary status of the
system using the system parameters with the latest evolutionary models, and to investigate the apsidal motion of the orbit for estimation of the
internal structures of the component stars.

\section{Observations}

\subsection{Photometry}

As the photometric data presented by B07 are precise enough (rms scatter of $\sim$0.008 mag) for light curve analysis with nearly complete phase
coverage of the data, apart from new times of minima observations, no new photometric observations were performed in this study. The details of
the photometric observations and their reduction are given in detail by B07. The standard {\em UBVR$_c$} and {\em I$_c$} magnitudes of the
comparison and check stars in the same CCD field with V731~Cep were derived using transformation coefficients obtained by B07, and are given
in Table~1 together with near-infrared magnitudes from the Two Micron Sky Survey ({\em 2MASS}, Cutri et al. 2003). {\em 2MASS} magnitudes in
Table~1 correspond to phases where the light was at maximum.

\begin{table*}
\small \centering \caption{Standard magnitudes for V731~Cep and the comparison stars. Infrared {\em J, H} and {\em K$_s$} magnitudes are taken
from {\em 2MASS} point sources catalogue (Cutri et al. 2003).} \label{table1}
\begin{tabular}{lccccccccc}\hline\hline
GSC No      & Phase& $U$     & $B$     & $V$     & $R_c$ & $I_c$ &$J$        &$H$         & $K_s$    \\
\hline
V731~Cep    & 0.00 & --    & 11.38 & 11.25 & 11.14 & 11.11 & -       &  -       &   -      \\
(4288 0186) & 0.25 & 10.59 & 10.63 & 10.54 & 10.46 & 10.42 & 10.095  &  10.049  &  10.021  \\
            & 0.51 & --    & 11.07 & 11.02 & 10.94 & 10.91 & -       &  -       &   -      \\
4288 0052   & --   & 10.95 & 10.95 & 10.86 & 10.76 & 10.70 & 10.329  &  10.297  &  10.252  \\
4288 0062   & --   & 11.57 & 11.42 & 11.14 & 10.88 & 10.70 & -       &  -       &  -       \\
4288 0220   & --   & 13.02 & 12.34 & 11.33 & 10.72 & 10.24 &  9.112  &   8.593  &   8.475  \\
4288 0241   & --   & 12.65 & 11.97 & 11.53 & 11.23 & 11.02 & 10.428  &  10.198  &  10.123  \\
\hline
\end{tabular}
\end{table*}

The phasing of the photometric data in the present paper was made using the updated ephemeris for primary minimum given in Eq. 1. This
ephemeris together with the ephemeris for the secondary minimum are given in B07. In Fig.~1, the primary and secondary minima of the light curve
is plotted together with $(B-V)$ colours.

\begin{equation}
Min I (HJD) = 2453137.4425(2) + 6.068456(2) \times E
\end{equation}

\begin{figure}
\begin{center}
\begin{tabular}{c}
    \resizebox{80mm}{!}{\includegraphics{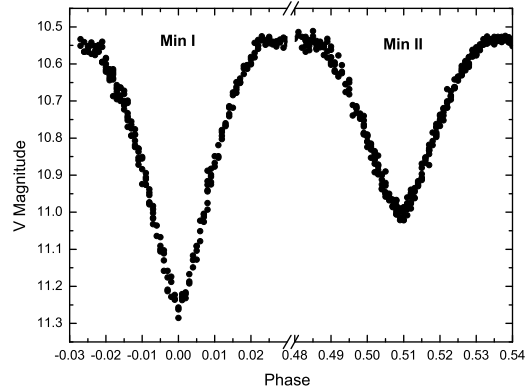}} \\
    \resizebox{80mm}{!}{\includegraphics{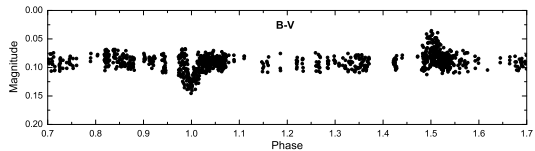}}
\end{tabular}
\caption{Primary and secondary minimum in {\em V}-band \textit{(upper panel)}. $B-V$ colour vs orbital phase \textit{(lower panel)}.}
\label{fig1}
\end{center}
\end{figure}

The difference between the periods given in Eq. 1 and in the ephemeris for the secondary minimum (B07) is clearly outside the observational
error box, which of course indicates an apsidal motion in the system. To study apsidal motion, new times of minima for V731~Cep were obtained
and are given in Table~2 together with the telescope and observatory information. During the (additional) eclipse timing observations at
Ond\v{r}ejov Observatory, one of the two available CCD cameras (SBIG ST-8 or Apogee AP-7) were used, while SBIG ST-10XME and ST-7 CCD cameras
were employed at \c{C}OM\"{U} and Brno Observatories, respectively. The reduction of all photometric data was performed using
C-Munipack\footnote{http://integral.physics.muni.cz/cmunipack/} free software under GNU General Public License. The times of minima have been
calculated according to the Kwee--van Woerden method (Kwee \& van Woerden 1956). In addition to new times of minima observations, all available
historical eclipse timings are also listed in Table~2.

\begin{table*}
\begin{center}
\caption{New and historical times of minima of V731~Cep.\label{table2}}
\begin{tabular}{lccccl}
\hline \hline
HJD         & Uncertainty & Weight    &    Filter  &  Type    & Reference / \\
(-2400000)  &             &           &            &          & Telescope and Observatory \\
\hline
52515.4840  & 0.0045  &  3        &clear         &  sec     & B03\\
52560.9401  & 0.0006  &  10       &$BVR_cI_c$    &  pri     & Lee et al. (2007) \\
52591.2817  & 0.0001  &  10       &clear         &  pri     & B03\\
52594.3767  & 0.0014  &  3        &clear         &  sec     & B03\\
52791.5388  & 0.0003  &  3        &clear         &  pri     & Bak{\i}\c{s} et al. (2005a)\\
52861.3828  & 0.0003  &  10       &clear         &  sec     & Bak{\i}\c{s} et al. (2005a)\\
52864.3624  & 0.0006  &  10       &clear         &  pri     & Bak{\i}\c{s} et al. (2005a)\\
52864.3626  & 0.0019  &  5        &$I$           &  pri     & Zejda (2004)\\
52864.3621  & 0.0028  &  5        &$V$           &  pri     & Zejda (2004)\\
52864.3630  & 0.0028  &  5        &$R$           &  pri     & Zejda (2004)\\
52867.4522  & 0.0014  &  5        &clear         &  sec     & Bak{\i}\c{s} et al. (2005a)\\
52867.4505  & 0.0013  &  10       &$RI$          &  sec     & Pejcha (2005)\\
52879.5890  & 0.0012  &  10       &clear         &  sec     & Bak{\i}\c{s} et al. (2005a)\\
52934.2099  & 0.0005  &  3        &clear         &  sec     & Bak{\i}\c{s} et al. (2005a)\\
52949.3211  & 0.0002  &  10       &$R_c$         &  pri     & 0.65 m reflector, primary focus, Ond\v{r}ejov Obs.\\
52967.5258  & 0.0004  &  10       &clear         &  pri     & Bak{\i}\c{s} et al. (2005a)\\
53134.4590  & 0.0004  &  3        &$BVR_cI_c$    &  sec     & Bak{\i}\c{s} et al. (2005a)\\
53137.4431  & 0.0005  &  10       &$BVI_c$       &  pri     & Bak{\i}\c{s} et al. (2005a)\\
53213.3532  & 0.0003  &  10       &$BV$          &  sec     & Bak{\i}\c{s} et al. (2005a)\\
53219.4218  & 0.0004  &  10       &$BVR_cI_c$    &  sec     & Bak{\i}\c{s} et al. (2005a)\\
53577.4608  & 0.0003  &  10       &$BVR_cI_c$    &  sec     & Bak{\i}\c{s} et al. (2005b)\\
53580.4393  & 0.0004  &  10       &$R_c$         &  pri     & 0.65 m reflector, primary focus, Ond\v{r}ejov Obs.\\
53580.4396  & 0.0002  &  10       &$BVR_cI_c$    &  pri     & Bak{\i}\c{s} et al. (2005b)\\
53753.4455  & 0.0003  &  10       &$BVR_cI_c$    &  sec     & B07\\
53756.4251  & 0.0003  &  10       &$BVR_cI_c$    &  pri     & B07\\
53929.4303  & 0.0004  &  10       &$R_c$         &  sec     & 0.65 m reflector, primary focus, Ond\v{r}ejov Obs.\\
53941.5675  & 0.0002  &  10       &$VR_c$        &  sec     & 0.40 m, Newton, Brno Obs.  \\
53944.54653 & 0.00012 &  20       &$R_c$         &  pri     & 0.65 m reflector, primary focus, Ond\v{r}ejov Obs.\\
54378.4960  & 0.0010  &  3        &clear         &  sec     & 0.30 m, Cassegrain-Schmidt, \c{C}OM\"{U} Obs.\\
54387.54272 & 0.00008 &  10       &$VR_c$        &  pri     & 0.40 m, Newton, Brno Obs.\\
54536.2750  & 0.0006  &  10       &$VR_cI_c$     &  sec     & 0.30 m, Cassegrain-Schmidt, \c{C}OM\"{U} Obs.\\
\hline
\end{tabular}
\end{center}
\end{table*}

\subsection{Spectroscopy}

Spectroscopic observations of V731~Cep were made using a Coud\'e spectrograph with 2.0-m reflector at Ond\v{r}ejov Observatory, equipped with a SITe-005
800$\times$2000 CCD. The spectra cover the region from 6280 \AA\,\,to 6720 \AA, with a linear dispersion of 17 \AA/mm and two-pixel resolution
of 12700. Each frame was processed using \textsc{IRAF}\footnote{IRAF is distributed by the National Optical Astronomy Observatories,
which is operated by the Association of Universities for Research in Astronomy, Inc. (AURA) under cooperative agreement with the National
Science Foundation.} software according to normal procedures of bias and dark substraction, flat-field division, rectification of background
continuum and wavelength calibration.

A total of seven spectra were obtained between 2005-2006. The exposure times of a single observation were arranged not to exceed 8000s which
corresponds to 0.015 in phase and to have an average signal-to-noise (S/N) ratio of $\sim$100 in continuum near H$_\alpha$ line. Two spectra
taken have S/N ratio below 100 due to bad weather conditions. The information on each individual spectrum is given in the journal of
spectroscopic observations presented in Table~3.

\begin{table*}
\begin{center}
\caption{Journal of spectroscopic observations. Date column refers
to the local date of the start of night. Signal-to-noise S/N ratio
refers to the continuum near H$_\alpha$.\label{table3}}
\begin{tabular}{ccccccc}
\hline \hline
No  &   Frame & HJD         &   Date    &   Phase    &   S/N      &  Exp. Time   \\
    &         & (-2400000)  &           &  ($\phi$)  &            &    (s)       \\
\hline
1   &  oj090013.fit  & 53653.32491  & 09.10.2005 & 0.010 &  50 &  7500 \\
2   &  pi090066.fit  & 53988.61624  & 10.09.2006 & 0.262 &  65 &  7000 \\
3   &  oh290012.fit  & 53612.44057  & 29.08.2005 & 0.273 & 115 &  7500 \\
4   &  pi100032.fit  & 53989.50350  & 11.09.2006 & 0.408 & 100 &  7200 \\
5   &  oh300036.fit  & 53613.47736  & 30.08.2005 & 0.444 & 100 &  8000 \\
6   &  oi240017.fit  & 53638.33066  & 24.09.2005 & 0.540 & 125 &  7200 \\
7   &  oi070028.fit  & 53621.42400  & 07.09.2005 & 0.754 & 115 &  7200 \\
\hline
\end{tabular}
\end{center}
\end{table*}

\section{Apsidal Motion}

The very slow apsidal motion of V731~Cep was detected and studied by
means of the O--C diagram analysis. All available times of minimum
were collected from the literature and are listed in Table~2
together with the new times of minima. We used the methods of
Gim\'enez \& Garc\'{\i}a-Pelayo (1983) and Lacy (1992) independently
with similar results given in Table~4. All precise CCD times of
minima were used with a weight of 20 or 10, less accurate
measurements were assigned with weights of 5 or 3.

Since the eccentricity is only weakly constrained by the observations in this case, we adopted a value of $\textit{e}$ = 0.0165 from the
light-curve solutions given in Table~7. For the inclination angle we used $\textit{i}$ = 88$^\circ$.7, from the light curve solutions. The
results are given in Table~4. In this table, (\textit{P}$_{\rm{a}}$) and (\textit{P}$_{\rm{s}}$) are the anomalous and sidereal periods,
respectively. The zero epoch is given (\textit{T}$_{\rm{0}}$) and the corresponding position of the periastron is represented by
(\textit{$\omega$}$_{\rm{0}}$). The apsidal motion rate ($\dot{\omega}$) that we obtained seems to be statistically significant, $\dot{\omega}$
= 0.00060 (0.00015) $^\circ$ cycle$^{-1}$. This corresponds to an apsidal period of $\textit{U}$ = 10000(2500) yr. The ephemeris curve is shown
in Fig.~2, along with the residuals from the observed primary and secondary minima.

We tested the stability of the results with respect to our arbitrarily chosen weighting scheme. It turned out that the resulting parameters show
some dependence on the weighting. For this reason, as well as for other reasons discussed below, the results must be considered preliminary and
less certain.

\begin{table}
\caption {Apsidal motion elements of V731~Cep. Errors in brackets denote the last digits.} \label{table4}
\begin{center}
\begin{tabular}{lcc}
\noalign{\smallskip} \hline \noalign{\smallskip}
 Parameter             & Unit     & Value         \\
\noalign{\smallskip} \hline \noalign{\smallskip}
\textit{T}$_{0}$       &HJD       &2453137.44241(5)\\
\textit{P}$_{\rm{s}}$  &d         &6.0684499(35) \\
\textit{P}$_{\rm{a}}$  &d         &6.0684600(25) \\
$\textit{e}$           &          &0.0165                    \\
$\dot{\omega}$         &$^\circ$/cycle&0.00060(15) \\
$\omega_{0}$           &$^\circ$      &29.87(42)      \\
$\textit{U}$           &yr        &10000(2500)           \\
\hline
\end{tabular}
\end{center}
\end{table}

\begin{figure}
\begin{center}
\resizebox{90mm}{!}{\includegraphics{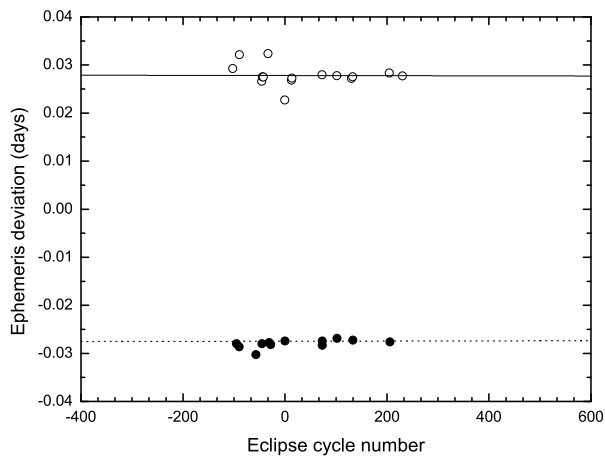}} \caption{Ephemeris curve for V731~Cep obtained with the parameters in Table~4. Filled and open
symbols represent the residuals from the primary and secondary minimum, respectively.} \label{fig2} \end{center}
\end{figure}

\section{Radial Velocities and Spectroscopic Orbit}

The method we adopted for radial velocity (RV) measurements was the two-dimensional cross-correlation ({\sc TODCOR}) of the observed spectra with
two synthetically produced template spectra. The algorithm of {\sc TODCOR} was developed by Zucker \& Mazeh (1994) and has been efficiently
applied to multiple-component spectra of late-type double-lined spectroscopic binaries (Latham et al. 1996; Metcalfe et al. 1996, among others)
and more recently to short-period early-type binaries (Gonzales \& Lapasset 2003; Southworth \& Clausen 2007). Basically, {\sc TODCOR}
calculates two-dimensional cross-correlation function (CCF) from one observed and two template spectra in re-binned log $\lambda$ space and then
locates the maximum of the CCF. We re-binned our observed and template spectra so that each pixel corresponded to 2.2 km$s^{-1}$. The
template spectrum of each component was synthetically produced using the appropriate model atmosphere grids of Kurucz (1993) for each component's
temperature, surface gravity and projected rotational velocity values, as listed in Table~8. The two template spectra are shown in Fig.~3 together
with the Si II doublets (6347.091 \AA, 6371.359 \AA) which were used for RV measurement by {\sc TODCOR}. The RVs of the components were
computed from the 2-dimensional CCFs with the highest score which is formed by combining the one dimensional CCFs of the shifted and rescaled
template spectra with the observed one. The RVs measured and their errors are presented in Table~5.

Using the {\sc TODCOR} RVs, the orbital solution was performed by least-squares orbital fitting. The orbital period of $P$$=$6.068456 days (Eq.
1) was fixed while the velocity semi-amplitudes $K_{1,2}$, systemic velocity $V_{\gamma}$ and the conjunction time (starting from 2453137.4425
HJD) were converged in the least-squares solution. Due to the small number of observations, the eccentricity ($e$) and the longitude of
periastron ($\omega$) parameters were also fixed at the values obtained from the light curve analysis results listed in Table~7. The radial
velocity at phase $\phi$$=$0.01 was not used due to inaccurate RV reading from the blended spectra of the components. In order to avoid
systematic effects in the orbital fitting due to the proximity of the components, although they are small, the effect of proximity was also
considered using the system geometry obtained in \S5.2. The orbital parameters adopted from the least-square fitting are listed in Table~6 and
the orbital fitting to RVs is shown in Fig.~4.

\begin{figure}
\begin{center}
\resizebox{90mm}{!}{\includegraphics[]{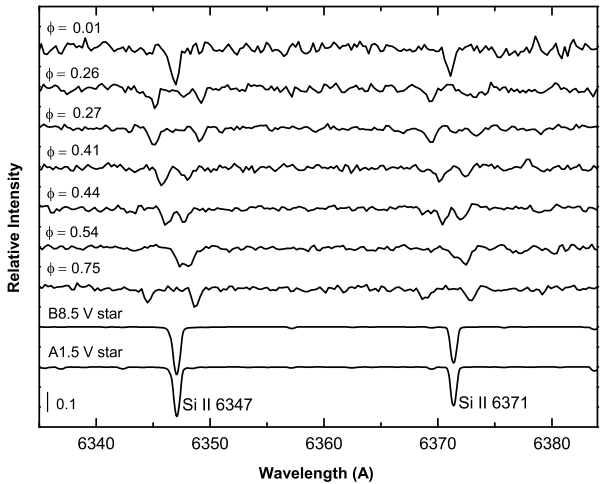}} \caption{Si II doublets (6347.091 \AA, 6371.359 \AA) at different orbital phases indicated.
Synthetic spectra of the components are shown below the observations and indicated with ``B8.5 V star'' for primary and ``A1.5 V star'' for
secondary components. For clarity, the intensity has been shifted and wavelength re-scaled.} \label{fig3} \end{center}
\end{figure}

\begin{table}
\begin{center}
\caption{RVs of the components measured by {\sc TODCOR}.\label{table5}}
\begin{tabular}{ccrr}\hline\hline
Time    &   Phase   & \multicolumn{1}{c}{$RV_{1}$}&\multicolumn{1}{c}{$RV_{2}$}\\
HJD     &   $\phi$  & \multicolumn{1}{c}{(km s$^{-1}$)}&\multicolumn{1}{c}{(km s$^{-1}$)}\\
\hline \\
2453653.32491 & 0.010 &   \multicolumn{1}{c}{---} & -5.7(1.2) \\
2453988.61624 & 0.262 &   -83.8(2.0)  & 107.3(3.2) \\
2453612.44057 & 0.273 &   -81.4(1.2)  & 109.3(1.8) \\
2453989.50350 & 0.408 &   -47.0(1.4)  & 57.1(2.1) \\
2453613.47736 & 0.444 &   -27.7(1.2)  & 43.8(1.4) \\
2453638.33066 & 0.540 &   14.2(1.3)   & -28.0(1.7) \\
2453621.42400 & 0.754 &   88.2(1.2)   & -106.1(1.5) \\
\hline
\end{tabular}
\end{center}
\end{table}

\begin{table}
\begin{center}
\caption{Spectroscopic orbital solution adopted from the analysis of {\sc TODCOR} RVs. Parameters without errors were fixed during the
solutions.\label{table6}}
\begin{tabular}{lc}\hline\hline
Parameter                  &   Value  \\
\hline
$P$ (d)                    &        6.068456 \\
T$_0$ (HJD-2453137)        & 0.322(0.030) \\
K$_{1}$ (km s$^{-1}$)      & 85.18(1.72)    \\
K$_{2}$ (km s$^{-1}$)      & 108.84(1.73)   \\
$q$ (K$_1$/K$_2$)          &0.783(0.024)  \\
V${_\gamma}$ (km s$^{-1}$) &0.62(0.94)     \\
$e$                        &    0.0165       \\
$w$ ($^\circ$)             &    25           \\
m$_{1}$sin$^{3}i$ (M$_\odot$)&    2.575(0.098)  \\
m$_{2}$sin$^{3}i$ (M$_\odot$)&    2.015(0.084)  \\
$a\sin i$         (R$_\odot$)&    23.26(0.29)  \\
\hline
\end{tabular}
\end{center}
\end{table}

\begin{figure}
\begin{center}
\resizebox{90mm}{!}{\includegraphics[]{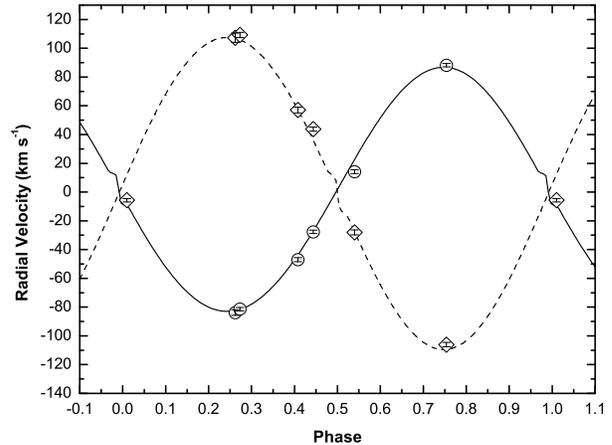}} \caption{RVs and the spectroscopic orbital solution. Primary and secondary RVs are shown with
circles and diamonds, respectively. Error of each individual RV measurement is indicated with bars in RV symbols.  Theoretical RV curves
including proximity and eclipse effects are shown for the primary (solid line) and secondary components (dashed line).} \label{fig4} \end{center}
\end{figure}

\section{Modelling of Light Curves}

\subsection{Reddening toward the system}

The first element needed in the light curve modelling is an estimate of the effective temperature of the primary component. The \emph{Q}-method
of Johnson \& Morgan (1953) was used with the {\em UBV} indices given in Table~1. Due to unavailability of {\em U}-magnitudes for eclipses, the
colours of the combined light ($\phi$$=$0.25) were used to determine a {\em Q} value of $-$0.10(0.03). According to the relation
$(B-V)_{0}$$=$$-$0.009$+$0.337\emph{Q} (Johnson \& Morgan 1953), one can determine the unreddened colour index of $(B-V)_{0}$=$-$0.04(0.02).
From the difference between the colours at 0.25 and 0.51 phases, the colour of the primary component is derived to be $(B-V)_{0}$=$-$0.07(0.02)
which corresponds to a spectral type of B8.5V with $E(B-V)$=0.13(0.03) colour excess.

The reddening obtained from the {\em Q}-method was checked by using Red Clump (RC) stars in this area. Red clump giants have long been proposed
as standard candles. They have a very narrow luminosity function and constitute a compact and well-defined clump in an HR diagram, particularly
in the infrared. Furthermore, as they are relatively luminous, they can be identified even at large distances from the Sun. The absolute
magnitude ($M_{K}$) and intrinsic colour, $(J-K_{s})_{0}$, of the red clump giants are well established (Alves 2000; Grocholski \& Sarajedini
2002; Salaris \& Girardi 2002; Pietrzynski et al. 2003). Here, we assume an absolute magnitude for the red clump population of -1.62(0.03) mag
and an intrinsic colour of $(J-K_{S})_{0}=0.7$ mag.

In order to make the reddening estimation in the direction of V731~Cep, the {\em J}, {\em H} and {\em K$_{s}$} magnitudes of the sources in the
1 square degree field, whose central galactic coordinates are $(l=115^{\circ}, b=2^{\circ})$, were obtained from {\em 2MASS} point sources
catalogue (Cutri et al. 2003). The isolation of the red clump sources in the star field was made by using theoretical traces to define the
limits of the K-giant branch on the colour magnitude diagrams (CMDs), without any further implication in the method -the traces were obtained for
different stellar types by using a double exponential approximation to the interstellar extinction according to the updated ``SKY'' model
(Wainscoat et al. 1992).

The maxima of the RC distribution, obtained via Gaussian fitting at different magnitude bins are shown in Fig.~5 (see Lopez-Corredoira et al.
2002, or Cabrera-Lavers et al. 2005 for further details of the method), and the distance of RC stars in the direction of V731~Cep and their
absorption on $K$-band are shown in Fig.~6. It is seen that the distance between RC stars in the direction of the star field and the Sun is
$1<d\leq7$ kpc. The average absorption in the $K$ band calculated from the RC stars closest ($1<d<1.5$ kpc) to V731~Cep system is
$A(K_{s})=0.279$. By using $A(K_{s})=0.382\times E(B-V)$ relation, the colour excess in the direction was calculated as $E(B-V)=0.730$ and total
absorption in $V$ band was calculated as $A(V)=2.263$. To reduce this colour excess to the V731~Cep system, we applied the following equation
(Bahcall \& Soneria 1980):

\begin{equation}
A_{d}(b)=A_{RC}(b)\Biggl[1-exp\Biggl(\frac{-\mid
d~sin(b)\mid}{H}\Biggr)\Biggr].
\end{equation}

\noindent Here, $b$ and $d$ are the Galactic latitude and distance of V731~Cep, respectively. $H$ is the scaleheight for the interstellar dust
adopted as 134 pc (Drimmel et al. 2003), $d=809$ pc, and $A_{RC}(b)$ and $A_{d}(b)$ are the total absorptions obtained from the SKY model for RC
stars and for the distance to the V731~Cep, respectively. According to Eq. 2, the total absorption in {\em V}-band and colour excess for the
distance to the V731~Cep are $A_{d}(b)=0.477(0.090)$ and $E(B-V)=0.154(0.029)$, respectively. The colour excess $E(B-V)=0.154(0.029)$ which is
found from RC stars is in good agreement with the $E(B-V)=0.13(0.03)$ obtained from the {\em Q}-method. Hence, it seems fair to adopt
confidently the unreddened colour of $(B-V)_0$$=$$-$0.07 mag as the intrinsic colour of the primary component. The intrinsic colour of the
primary component corresponds to a spectral type of B8.5V (Fitzgerald 1970) with a temperature of 10700 K according to the calibration tables of
Strai\v{z}ys \& Kuriliene (1981).

\begin{figure}
\begin{center}
\resizebox{80mm}{!}{\includegraphics{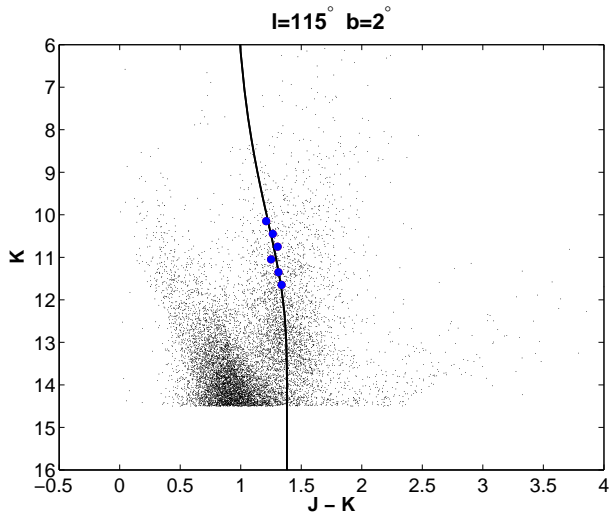}} \caption[] {Maxima of the RC stars for different magnitude bins and the theoretical trace for a
RC star (spectral type K2-3III) obtained from the SKY model by assuming a double  exponential distribution for the dust in the
Galaxy.}\label{fig5}
\end{center}
\end {figure}

\begin{figure}
\begin{center}
\resizebox{80mm}{!}{\includegraphics{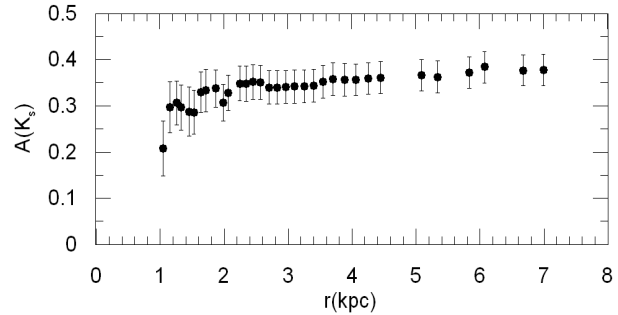}} \caption[]
{Extinction along the line of sight ($A_{K}$ vs. $r$) for the field
$(l=115^{\circ}, b=2^{\circ})$ obtained from the RC maxima in the
CMD.}\label{fig6}
\end{center}
\end {figure}

\subsection{Determination of the photometric elements}

We performed a simultaneous analysis of the {\em BVR$_c$} and {\em I$_c$} light curves using the 2003 version of the Wilson-Devinney (Wilson \&
Devinney 1971; Wilson 1994; hereafter WD) program.

During the simultaneous fit with WD code, the adjustable parameters were the orbital semi-major axis ($a$), the orbital eccentricity ($e$), the
longitude of periastron ($\omega$), the orbital inclination ($i$), the temperature of the secondary component ($T_{eff2}$), surface potential of
both components ($\Omega_{1,2}$), and the luminosity of the primary component in each band ($L_{1}$). The mass ratio ($q$) was taken from the
spectroscopic orbital solution (\S3) and was used as a fixed parameter during the analysis. An eccentric orbit was adopted, and the gravity
brightening coefficients and bolometric albedos were set to unity in accordance with the radiative atmospheres of the components. The
limb-darkening (LD) coefficients from the logarithmic LD-law were computed at each iteration from van Hamme (1993).

All observations in each band were weighted equally. Convergence of the fits was reached rapidly and tests from different starting points
indicated the uniqueness of the solution. The final residual sum of squares (rss) are similar in all bands of the light curves, 0.038 in {\em
B}, 0.040 in {\em V}, 0.033 in {\em R$_c$} and 0.042 in {\em I$_c$}. The resulting best fitting light curve elements are given in Table~7. The
light curves with the best fitting model curves superimposed and the O--C residuals from the fits are shown in Fig.~7.

\begin{table}
\begin{center}
\caption{Results from the simultaneous solution of {\em BVR$_c$} and {\em I$_c$}-band light curves of V731~Cep system. Adjusted and fixed
parameters are presented in separate panels of the table.} \label{table7}
\begin{tabular}{lc}\hline\hline
Parameter &  Value \\
\hline
\multicolumn{2}{l}{Adjusted parameters:}  \\
\hline
$T_{eff2}(K)$   &   9265(20) \\
$L_{1}/L_{1+2}(B)$    &   0.639(0.008) \\
$L_{1}/L_{1+2}(V)$    &   0.620(0.008) \\
$L_{1}/L_{1+2}(R_c)$    &   0.613(0.008) \\
$L_{1}/L_{1+2}(I_c)$    &   0.604(0.008) \\
$\Omega_1$   &  13.24(0.12)        \\
$\Omega_2$   &  11.75(0.08)        \\
$r_{1}$(mean)&   0.0805(0.0008) \\
$r_{2}$(mean)&   0.0738(0.0005) \\
$i (^{o})$   &   88.70(0.03)  \\
$e$          &   0.0165(0.0005) \\
$w (^{o})$   &   25(2)   \\
\hline
\multicolumn{2}{l}{Fixed parameters:}  \\
\hline
A$_1$=A$_2$ & 1.0  \\
g$_1$=g$_2$ & 1.0  \\
$T_{eff1}(K)$  &   10700    \\
$q$         &   0.783    \\
x$_1(B,V,R_c,I_c)$ & 0.678, 0.581, 0.496, 0.396  \\
y$_1(B,V,R_c,I_c)$ & 0.338, 0.290, 0.246, 0.195  \\
x$_2(B,V,R_c,I_c)$ & 0.733, 0.630, 0.534, 0.425  \\
y$_2(B,V,R_c,I_c)$ & 0.325, 0.286, 0.249, 0.201  \\
$F_{1}$=$F{_2}$ & \multicolumn{1}{c}{1.034} \\
\hline
$\chi^2_{min}$     &    0.149    \\
\hline
\end{tabular}
\end{center}
\end{table}

\begin{figure*}
\begin{center}
\begin{tabular}{c}
    \resizebox{120mm}{80mm}{\includegraphics{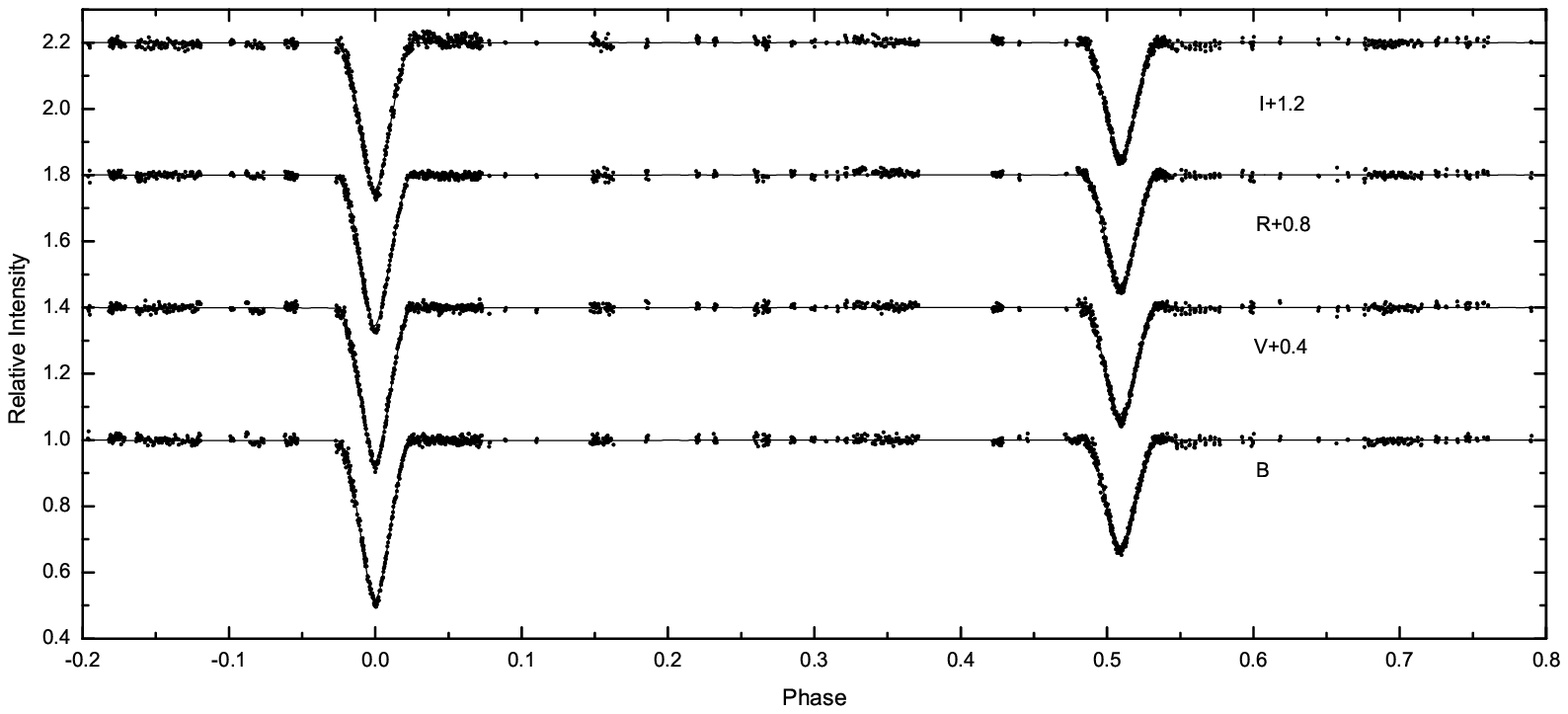}} \\
    \resizebox{100mm}{!}{\includegraphics{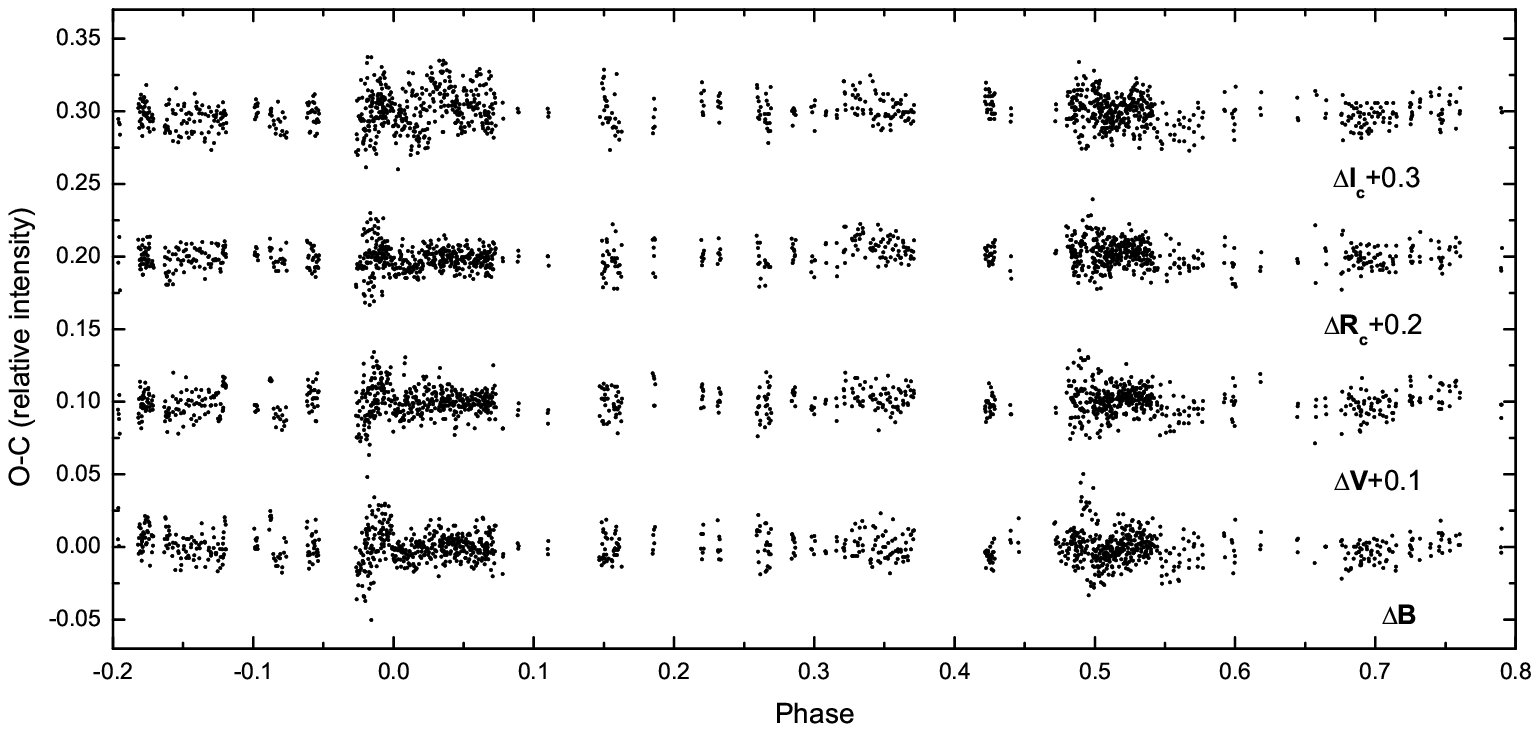}}
\end{tabular}
\caption{Theoretical model fits to the light curves of V731 Cep \textit{(upper panel)}. O--C residuals from the theoretical fits \textit{(lower
panel)}.}\label{fig7}
\end{center}
\end{figure*}

\section{Discussion}

\subsection{Absolute Dimensions and Distance of the System}

Combination of spectroscopic orbital elements (Table~6) with light curve elements (Table~7) yields the absolute dimensions of the system,
which are presented in Table~8. The adopted temperature T$_{eff1}$$=$10700 K and mass $M_{1}$$=$2.577$M_{\odot}$ of the primary component
correspond to the spectral type of a normal B8.5-type main sequence star, while the adopted temperature T$_{eff2}$$=$9265 K and mass
$M_{2}$$=$2.017$M_{\odot}$ of the secondary component are in good agreement with the spectral type of a normal A1.5-type main sequence star
(i.e. Strai\v{z}ys \& Kuriliene 1981). However, the radii of the components, $R_{1}$$=$1.823$R_{\odot}$ and $R_{2}$$=$1.717$R_{\odot}$, are in
better agreement with the same spectral type stars but at closer locations to ZAMS, suggesting a young age of the system, as determined in \S6.2.

The synchronization time-scale for the component stars of the V731~Cep system is, following Zahn (1977), in the order of 15 Myr, which is
smaller than the age of 120 Myr estimated from the isochrones (see \S6.2). To compare the observed rotational velocities with the
synchronization velocities listed in Table~8, we have modelled Si II doublets (6347.091 \AA, 6371.359 \AA) with model atmosphere grids using
ATLAS9 and SYNTHE codes under Linux (Kurucz 1993). The modeling yielded equatorial rotational velocities of V$_{rot1}$$=$19(3) km s$^{-1}$ and
V$_{rot2}$$=$18(3) km s$^{-1}$ for the primary and secondary components, respectively. Although errors in the observed rotational velocities are
in the order of 15 per cent, the synchronization velocities seem to be slightly below the observational measurements. The asynchronization of
the components with the orbit should be confirmed by analyzing new spectra with a higher S/N ratio and more absorption lines in a larger
spectral range.

\begin{table*}
\small \caption{Close binary stellar parameters of V731~Cep. Errors
of parameters are given in parenthesis.} \label{table8}
\begin{tabular}{lccc}\hline\hline
Parameter                          & Symbol  & Primary & Secondary  \\
\hline
Mass (M$_\odot$)                   & \emph{M}       & 2.577(0.098) & 2.017(0.084)    \\
Radius (R$_\odot$)                 & \emph{R}       & 1.823(0.030) & 1.717(0.025)    \\
Separation (R$_\odot$)             & \emph{a}     & \multicolumn{2}{c}{23.27(0.29)} \\
Surface gravity (cgs)              & log $g$        &4.304(0.011)& 4.273(0.011)   \\
Integrated visual magnitude (mag)  &  \emph{V}      &\multicolumn{2}{c}{10.54(0.01)}\\
Integrated colour index (mag)       & $B-V$          &\multicolumn{2}{c}{0.09(0.01)}\\
Colour excess (mag)                 &$E(B-V)$&\multicolumn{2}{c}{0.13(0.03)}\\
Visual absorption (mag)            & $A_{V}$ &\multicolumn{2}{c}{0.40}\\
Intrinsic colour index (mag)        & $(B-V)_{0}$&\multicolumn{2}{c}{-0.04(0.02)}\\
Component intrinsic colour index (mag) & $(B-V)$ & -0.073(0.020) & 0.016(0.020)  \\
Temperature (K)                    & $T_{eff}$ & 10700(200) & 9265(220) \\
Spectral type                      & Sp        & B8.5 V   & A1.5 V        \\
Luminosity (L$_\odot$)             & log \emph{L}& 1.618(0.035) & 1.292(0.043)\\
Computed synchronization velocities (km s$^{-1}$)& V$_{synch}$ & 15.6(0.2) & 14.3(0.2) \\
Observed rotational velocities (km s$^{-1}$) & V$_{rot}$ & 19(3) & 18(3) \\
Bolometric magnitude (mag)         &$M_{bol}$& 0.705(0.088) & 1.519(0.108)    \\
Velocity amplitudes (km s$^{-1}$)  &$K_{1,2}$& 85.18(1.72) & 108.84(1.73)    \\
Absolute visual magnitude (mag)    &$M_{v}$  & 1.104(0.054) & 1.631(0.070)    \\
Bolometric correction (mag)        &\emph{BC}& -0.399      & -0.112      \\
Distance (pc)                      &\emph{d} & \multicolumn{2}{c}{809(30)} \\
Systemic velocity (km s$^{-1}$)    &$V_{\gamma}$ & \multicolumn{2}{c}{0.62(0.94)} \\
Parallax (mas)                     &$\pi$        & \multicolumn{2}{c}{1.236(0.044)*} \\
Proper motion (mas yr$^{-1}$)      &$\mu_\alpha cos\delta$, $\mu_\delta$ & \multicolumn{2}{c}{-1.5 (2.6), -3.1 (2.5)**} \\
Space velocities (km s$^{-1}$)     & $U, V, W$  & \multicolumn{2}{c}{7.59(9.03), 4.71(4.31), -9.75(9.62)}\\
\hline
{\em * In this study.} \\
{\em ** NOMAD Catalog (Zacharias 2005).}
\end{tabular}
\end{table*}

Using the brightness of the system listed in Table~1 together with the light contributions of the components listed in Table~7, the intrinsic
magnitudes of the components were calculated and are presented in Table~9. During the derivation of the intrinsic magnitudes, the interstellar
extinction in {\em B} and {\em V} bands were adopted from the $Q$-method while the following relations of Fiorucci \& Munari (2003) were used for
the determination of extinction in $R_c$ and $I_c$.

\begin{eqnarray}
\nonumber (R_{c})_0 = R_c - 2.494 \times E(B-V), \\
(I_{c})_0 = I_c - 1.753\times E(B-V).
\end{eqnarray}

\noindent where ($R{_c})_0$ and ($I{_c})_0$ are the de-reddened magnitudes.

\begin{table}
\begin{center}
\caption{De-reddened magnitudes of stars in V731~Cep system.} \label{table9}
\begin{tabular}{lccccc}\hline\hline
         & {\em B}     & {\em V}     & {\em R$_c$} & {\em I$_c$} & Err.\\
\hline
{\em Primary}   & 10.59 & 10.66 & 10.67 & 10.74 & 0.02 \\
{\em Secondary} & 11.21 & 11.19 & 11.17 & 11.20 & 0.02 \\
\hline
\end{tabular}
\end{center}
\end{table}

The de-reddened visual apparent magnitude and optical absolute magnitude presented in Table~8 allowed us to derive a distance of 809(30) pc to
the system. To compare the distance of V731~Cep system using a different method, a luminosity-colour relation (Bilir et al. 2008) which has been
formed for detached binary systems with main-sequence components was used in this study. The near-infrared magnitudes of the system were taken
from the {\em 2MASS} Point Sources Catalogue of Cutri et al. (2003) and are shown in Table~1. For de-reddenig near-infrared magnitude and
colours of the system, the following formulae (Bilir, G\"uver, Aslan 2006; Ak et al. 2007; Bilir et al. 2008) were used:

\begin{eqnarray}
\nonumber J_{o}=J-0.884\times E(B-V), \\
(J-H)_{o}=(J-H)-0.322\times E(B-V), \\
\nonumber (H-K_{s})_{o}=(H-K_{s})-0.187\times E(B-V).
\end{eqnarray}

All the colours and magnitude with subscript ``0'' show the de-reddened ones. The colour excess $E(B-V)=0.13$  was estimated in a direction to
V731~Cep by using {\em Q}-method (see \S5.1). The near-infrared absolute magnitude of V731~Cep system was estimated by the luminosity-colour
relation, $M_{J}=5.228(J-H)_{o}+6.185(H-K_{s})_{0}+0.608$, of Bilir et al. (2008) and the distance of the system calculated as $733(50)$ pc by
using the photometric parallaxes method. The photometric distance of $809(30)$ pc given in Table~8 is consistent with the $733(50)$ pc distance
estimated by luminosity-colour relation obtained for detached binary systems.

\subsection{Internal Structure}

Binary systems with apsidal motion allow us to determine the ISC,
which is an important parameter of stellar evolution models. The
observed apsidal motion period of $\textit{U}$ = 10000(2500) yr,
corresponding to a total rate of $\dot{\omega}$ = 0.00060(0.00015)
$^\circ$ cycle$^{-1}$ was obtained in \S6. The relativistic
contribution to the apsidal motion in case of V731~Cep is
substantial $\dot{\omega}_{rel}$ = 0.00045 $^\circ$ cycle$^{-1}$, or
about 75 per cent of the total observed rate (Gim\'enez 1985). After
correcting for this effect, an average ISC was derived to be log
$\bar{k}_{2,obs}$ = $-$2.36 under the assumption that the component
stars rotate pseudosynchronously. This value is in very good
agreement with the theoretical prediction of log $\bar{k}_{2,theo}$
= $-$2.34 according to new evolutionary models of Claret (2004) with
the standard chemical composition of (X,Z) = (0.70, 0.02). It
should, however, be noted that the present apsidal motion solution
is still tentative due to relatively short observational history of
V731 Cep compared to the apsidal motion period. Therefore, accurate
eclipse timings are strongly needed in a decade or more in order to
say more definite on the apsidal motion parameters and the related
ISC.

\subsection{Evolutionary Stage and Age of the System}

We investigated the evolutionary status of the system by means of
the Bayesian method and constructing an H-R diagram for the
component masses in log T$_{eff}$-log L plane. We used a slightly
modified version of the Bayesian estimation method idealized by
J{\o}rgensen \& Lindegren (2005), which is designed to avoid
statistical biases and to take error estimates of all observed
quantities into consideration. Estimation of age and metal abundance
of the components was made by using the web
interface\footnote{http://stev.oapd.inaf.it/$\sim$lgirardi/cgi-bin/param}
based on the Bayesian method of da Silva et al. (2006). Including
their errors, the effective temperatures, visual brightness, metal
abundance of components and distance to the system were the
parameters used in the web interface to obtain the best matching
model parameters (i.e. surface gravity (log $g$), radii ($R_{1,2}$),
masses ($M_{1,2}$) and age of components ($t_{1,2}$)) to Padova
isochrones by the Bayesian method. Since the metal abundance of the
components was initially not known, a range of metal abundance (i.e.
$-$0.10 $<$ $[M/H]$ $<$ +0.10 dex) was selected and the values in
this range were used with 0.02 dex steps. The output model
parameters for each component were compared by means of $\chi^2$
test with the absolute dimensions of the components listed in
Table~8. Consequently, the minimum $\chi^2$ yielded simultaneously
the metal abundance and age of the components as
$[M/H]_1=-0.06(0.02)$ dex and $t_{1}=116(15)$ Myr for the primary
and $[M/H]_2=-0.02(0.02)$ dex and $t_{2}=150(15)$ Myr for the
secondary component. From these values, we adopted the mean metal
abundance and mean age of the system to be $[M/H]=-0.04(0.02)$ dex
and $t=133(26)$ Myr, respectively.

Interpretation of the evolutionary status of V731~Cep requires also
the construction of H-R diagrams using the latest theoretical
evolutionary models. In Fig.~8, the components of V731~Cep are shown
in the log T$_{eff}$-log $L$ plane together with Yonsei-Yale (Y2)
evolutionary tracks (Yi, Demarque, Kim et al. 2001; hereafter YDK)
for different masses. Evolutionary tracks for the exact masses of
the components were also computed using the code provided by YDK.
Among the tracks computed for the exact masses with their errors,
those with $[M/H]$$=$-0.024 dex ($Z$$=$0.0172) metallicity models
match the locations of the components within the error limits in the
log T$_{eff}$-log $L$ plane. We also computed a set of isochrones
using a metallicity of $[M/H]$$=$ $-$0.024 dex in Y2 models. In
Fig.~8, two isochrones ($t=$100 Myr and $t=$120 Myr) are plotted for
comparison. It was found that 120 Myr age is the best fitting
isochrone to the locations of both components. Although, the most
reliable method of metal abundance determination is the atmosphere
modeling of spectral lines, in the present study, it seems fair to
conclude that the two methods (Bayesian methods and evolutionary
tracks) used for estimation of the metallicity and age of the system
are in excellent agreement within the error limits. The metallicity
we found for the components should be confirmed with the atmosphere
modeling of metallic absorption lines in spectra taken in wider
wavelength range.

\begin{figure}
\centering \resizebox{90mm}{!} {\includegraphics[]{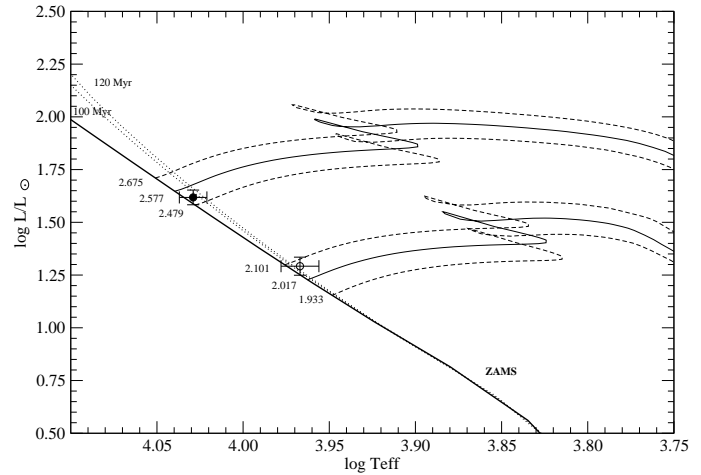}} \caption{Evolutionary tracks for individual component masses and isochrone curves
best matching the location of the components in log T$_{eff}$ - log $L$ plane. The primary and secondary stars are shown with filled and empty
circles, respectively.} \label{fig8}
\end{figure}

\subsection{Possible Origin of V731~Cep}

One of the formation regions of early-type stars is open clusters. To find a possible formation region for V731~Cep, nearby young open clusters
were investigated. Among others, NGC~7762 was found to be the closest open cluster and the most similar in chemical composition to V731~Cep.
Chincarini (1966) estimated the distance of NGC~7762 to be 750 pc and the age 266 Myr. In a more recent study relating on the cluster, Patat \&
Carraro (1995) proposed a similar distance of 800 pc to NGC~7762. Nevertheless, Patat \& Carraro (1995) estimated an older age for the cluster
at 1.8 Gyr and less metallicity compared to the Sun. The age (133 Myr) we adopted for V731~Cep system seems to agree more with the age (266 Myr)
that Chincarini (1966) calculated for NGC~7762. In addition to this, the distance and metal abundance estimation of NGC~7762 by Patat \& Carraro
(1995) are in agreement with the distance and metal abundance of V731~Cep found in this study. The most reliable evidence that V731~Cep is
evaporated from NGC~7762 can be obtained from the spatial distribution and the metallicity of V731~Cep.

In calculation of the kinematical properties of V731~Cep, the systemic velocity, distance and proper motion components listed in Table~8
were used in the algorithm given by Johnson \& Soderbloom (1987). The computed space velocity components with their errors are given in Table~8.
The total space velocity of 14 km s$^{-1}$ for V731~Cep is in agreement with the space velocities of young stars. Although the space velocity of
the cluster could not be computed due to its unavailable RV data, the distance of 55 pc between V731~Cep and NGC~7762, and similar age and metallicity
distribution of V731~Cep with NGC~7762, suggest that V731~Cep could be evaporated from NGC~7762. However, it is necessary to carry out precise
photometric and spectroscopic observations of the member stars of NGC~7762 in order to form more definite conclusions on the
history of V731~Cep in relation with NGC~7762. \\ \\

\textbf{Acknowledgements} \\
The authors would like to thank to Dr. Antonio Cabrera-Lavers for his help in extracting extinction data from RC stars and to the referee, Prof.
E. F. Guinan, for his useful comments that improved the readability of this paper. This study is part of a project funded by \c{C}OMU Scientific
Research Foundation under project code BAP2008/37. The research of MW was supported by the Research Program MSM0021620860 of the Ministry of
Education of Czech Republic. Participation of MZ in this study was endorsed by the grant
GA CR 205/06/0217 of the Czech Science Foundation. \\

\label{lastpage}


\begin{thebibliography}{}
\bibitem[\protect\citeauthoryear{Ak et al.}{2007}]{A1}Ak T., Bilir S., Ak S., Retter A., 2007, NewA, 12, 446
\bibitem[\protect\citeauthoryear{Alves}{2000}]{A2}Alves D. R., 2000, ApJ, 539, 732
\bibitem[\protect\citeauthoryear{Andersen}{1991}]{A3} Andersen J., 1991, A\&ARv, 3, 91
\bibitem[\protect\citeauthoryear{Bahcall \& Soneira}{1980}]{B0}Bahcall J.N., Soneira R.M., 1980, ApJS, 44, 73
\bibitem[\protect\citeauthoryear{Bakis}{2003}]{B1} Bak{\i}\c{s} V., Erdem A., Budding E., Demircan O., 2003, IBVS, 5381
\bibitem[\protect\citeauthoryear{Bakis}{2003}]{B2} Bak{\i}\c{s} V., Bak{\i}\c{s} H., T\"{u}ys\"{u}z M., \"{O}zkarde\c{s} B., Erdem A., \c{C}i\c{c}ek C., Demircan O., 2005a, IBVS, 5616
\bibitem[\protect\citeauthoryear{Bakis}{2003}]{B3} Bak{\i}\c{s} V., Do\v{g}ru S.S., Bak{\i}\c{s} H., Do\v{g}ru D., Erdem A., \c{C}i\c{c}ek C., Demircan O., 2005b, IBVS,5662
\bibitem[\protect\citeauthoryear{Bakis}{2007}]{B4} Bak{\i}\c{s} V., Bak{\i}\c{s} H., Budding E., Demircan O., Zejda M., 2007, in Solar and Stellar
Physics Through Eclipses (ASP Conf. Ser.: San Francisco), eds. O. Demircan, S.O. Selam and B. Albayrak, 370, 213
\bibitem[\protect\citeauthoryear{Bakis}{2008}]{B5} Bak{\i}\c{s} V., Bak{\i}\c{s} H., Demircan O., Eker Z., 2008, MNRAS, 384, 1657
\bibitem[\protect\citeauthoryear{Bilir, G\"uver \& Aslan}{2006}]{B4}Bilir S., G\"uver T., Aslan M., 2006, AN, 327, 693
\bibitem[\protect\citeauthoryear{Bilir et al.}{2008}]{B5}Bilir S., Ak T., Soydugan E., Soydugan F., Yaz E., Filiz Ak N., Eker Z.,
Demircan O., Helvac{\i} M., 2008, AN (in press) (astro/ph:
2008arXiv0806.1290)
\bibitem[\protect\citeauthoryear{Cabrera-Lavers et al.}{2005}]{C1}Cabrera-Lavers A., Garz{\' o}n F., Hammersley P. L., 2005, A\&A, 433, 173
\bibitem[\protect\citeauthoryear{Chincarini}{1966}]{C2} Chincarini G., 1966, MmSAI, 37, 423
\bibitem[\protect\citeauthoryear{Claret}{2004}]{C3} Claret A., 2004, A\&A, 424, 919
\bibitem[\protect\citeauthoryear{Cutri et al.}{2003}]{C4} Cutri R. M., et al., 2003, The IRSA 2MASS All-Sky Point Source Catalog,
NASA/IPAC Infrared Science Archive.~http://irsa.ipac.caltech.edu/applications/Gator/
\bibitem[\protect\citeauthoryear{da Silva et al.}{2006}]{D1}da Silva L., Girardi L., Pasquini L., Setiawan J., von der L\"uhe O., de
Medeiros J. R., Hatzes A., D\"ollinger M. P., Weiss A., 2006, A\&A, 458, 609
\bibitem[\protect\citeauthoryear{Drimmel, Cabrera-Lavers \& L\'opez-Corredoira}{2003}]{D2} Drimmel R., Cabrera-Lavers A.,
L\'opez-Corredoira M., 2003, A\&A, 409, 205
\bibitem[\protect\citeauthoryear{Fiorucci \& Munari}{2003}]{F1}Fiorucci M., Munari U., 2003, A\&A, 401, 781
\bibitem[\protect\citeauthoryear{Fitzgerald}{1970}]{F2} Fitzgerald M. P., 1970, A\&A, 4, 234
\bibitem[\protect\citeauthoryear{Gim\'enez}{1983}]{G1} Gim\'enez A., Garcia-Pelayo J.M., 1983, Ap\&SS, 92, 203
\bibitem[\protect\citeauthoryear{Gim\'enez}{1985}]{G12} Gim\'enez A., 1985, ApJ, 297, 405
\bibitem[\protect\citeauthoryear{Gonzales}{2003}]{G2} Gonz\'{a}lez J. F., Lapasset E., 2003, A\&A, 404, 365
\bibitem[\protect\citeauthoryear{Grocholski \& Sarajedini}{2002}]{G3}Grocholski A. J., Sarajedini A., 2002, AJ, 123, 1603
\bibitem[\protect\citeauthoryear{Johnson}{1953}]{J1} Johnson H. L., Morgan W. W., 1953, ApJ, 117, 313
\bibitem[\protect\citeauthoryear{Johnson \& Soderbloom}{1987}]{J2}Johnson D. R. H., Soderblom D. R., 1987, AJ, 93, 864
\bibitem[\protect\citeauthoryear{J{\o}rgensen \& Lindegren}{2005}]{J3}J{\o}rgensen B. R., Lindegren L., 2005, A\&A, 436, 127
\bibitem[\protect\citeauthoryear{Kurucz}{1993}]{K1} Kurucz R. L., 1993, CD-ROM 13, 18, http://kurucz.harward.edu
\bibitem[\protect\citeauthoryear{Kwee}{1956}]{K2} Kwee K. K., van Woerden, H., 1956, BAN, 12, 327
\bibitem[\protect\citeauthoryear{Lacy}{1992}]{L1} Lacy Claud H. S., 1992, AJ, 104, 2213
\bibitem[\protect\citeauthoryear{Latham}{1996}]{L2} Latham D. W., Nordstroem B., Andersen J., Torres G., Stefanik R. P.,
Thaller M., Bester M. J., 1996, A\&A, 314, 864
\bibitem[\protect\citeauthoryear{Lee}{2007}]{L3} Lee J. W., Kim C.-H., Koch R. H., 2007, MNRAS, 379, 1665
\bibitem[\protect\citeauthoryear{L\'opez-Corredoira et al.}{2002}]{L4}L\'opez-Corredoira M., Cabrera-Lavers A., Garz{\' o}n F., Hammersley P. L.,
2002, A\&A, 394, 883
\bibitem[\protect\citeauthoryear{Metcalfe}{1996}]{M1} Metcalfe T. S., Mathieu R. D., Latham D. W., Torres G., 1996, ApJ, 456, 356
\bibitem[\protect\citeauthoryear{Patat \& Carraro}{1995}]{P1}Patat F., Carraro G., 1995, A\&AS, 114, 281
\bibitem[\protect\citeauthoryear{Pejcha}{2005}]{P2} Pejcha O., 2005, IBVS, 5645
\bibitem[\protect\citeauthoryear{Pietrzy\'nski et al.}{2003}]{P3}Pietrzy{\'n}ski G., Gieren W., Udalski A., 2003, AJ, 125, 2494
\bibitem[\protect\citeauthoryear{Salaris \& Girardi}{2002}]{S1}Salaris M., Girardi L., 2002, MNRAS, 337, 332
\bibitem[\protect\citeauthoryear{Southworth}{2005}]{S2} Southworth J., Smalley B., Maxted P. F. L., Claret A., Etzel P. B., 2005, MNRAS, 363,
529
\bibitem[\protect\citeauthoryear{Southworth\&Clausen}{2007}]{S3} Southworth J., Clausen J. V., 2007, A\&A, 461, 1077
\bibitem[\protect\citeauthoryear{Straizys}{1981}]{S4} Strai\v{z}ys V., Kuriliene G., 1981, Ap\&SS, 80, 353
\bibitem[\protect\citeauthoryear{vanHamme}{1993}]{V1} van Hamme W., 1993, AJ, 106, 2096
\bibitem[\protect\citeauthoryear{Wainscoat et al.}{1992}]{W1}Wainscoat R. J., Cohen M., Volk K., Walzer H. J., Schwartz D. E., 1992,
ApJS, 83, 111
\bibitem[\protect\citeauthoryear{Wilson}{1971}]{W2} Wilson R. E., Devinney E. J., 1971, ApJ, 166, 605
\bibitem[\protect\citeauthoryear{Wilson}{1994}]{W3} Wilson R. E., 1994, PASP, 106, 921
\bibitem[\protect\citeauthoryear{Yi}{2001}]{Y1} Yi S., Demarque P., Kim Y., Lee Y., Ree C. H., Lejeune T., Barnes S., 2001, ApJS, 136, 417
\bibitem[\protect\citeauthoryear{Zucker}{2005}]{Z0} Zacharias N., Monet D.G., Levine S.E., Urban S.E., Gaume R., Wycoff G.L., 2005, Naval
Observatory Merged Astrometric Dataset (NOMAD), Vizier, http://cdsarc.u-strasbg.fr/viz-bin/Cat?I/297
\bibitem[\protect\citeauthoryear{Zahn}{1977}]{Z1} Zahn J.P., 1977, A\&A, 57, 383
\bibitem[\protect\citeauthoryear{Zejda}{2004}]{Z2} Zejda M., 2004, IBVS, 5583
\bibitem[\protect\citeauthoryear{Zucker}{1994}]{Z3} Zucker S., Mazeh T., 1994, ApJ, 420, 806
\end{thebibliography}
\end{document}